\documentclass[runningheads]{llncs}
\usepackage[caption=false]{subfig}
\usepackage{graphicx}
\usepackage{amsfonts}
\usepackage{mathtools}
\usepackage{bm}
\usepackage{multirow}
\DeclareMathOperator*{\argmax}{arg\,max}
\begin{document}
\title{Weakly Supervised Cell Instance Segmentation by Propagating from Detection Response}

%
\author{Kazuya Nishimura\inst{1} \and
Dai Fei Elmer Ker \inst{2}\and
Ryoma Bise\inst{1}}

\authorrunning{K. Nishimura et al.}
\titlerunning{Weakly Supervised Cell Instance Segmentation}

\institute{Kyushu University, Fukuoka City, Japan.
\email{kazuya.nishimura@human.ait.kyushu-u.ac.jp}
\and
The Chinese University of Hong Kong.}
\maketitle              
\begin{abstract}
Cell shape analysis is important in biomedical research.
Deep learning methods may perform to segment individual cells if they use sufficient training data that the boundary of each cell is annotated.
However, it is very time-consuming for preparing such detailed annotation for many cell culture conditions.
In this paper, we propose a weakly supervised method that can segment individual cell regions who touch each other with unclear boundaries in dense conditions without the training data for cell regions.
We demonstrated the efficacy of our method using several data-set including multiple cell types captured by several types of microscopy.
Our method achieved the highest accuracy compared with several conventional methods. In addition, we demonstrated that our method can perform without any annotation by using fluorescence images that cell nuclear were stained as training data.Code is publicly available in \url{https://github.com/naivete5656/WSISPDR}
\end{abstract}

\section{Introduction}
\begin{figure}[t]
    \centering
    \includegraphics[width=0.95\linewidth]{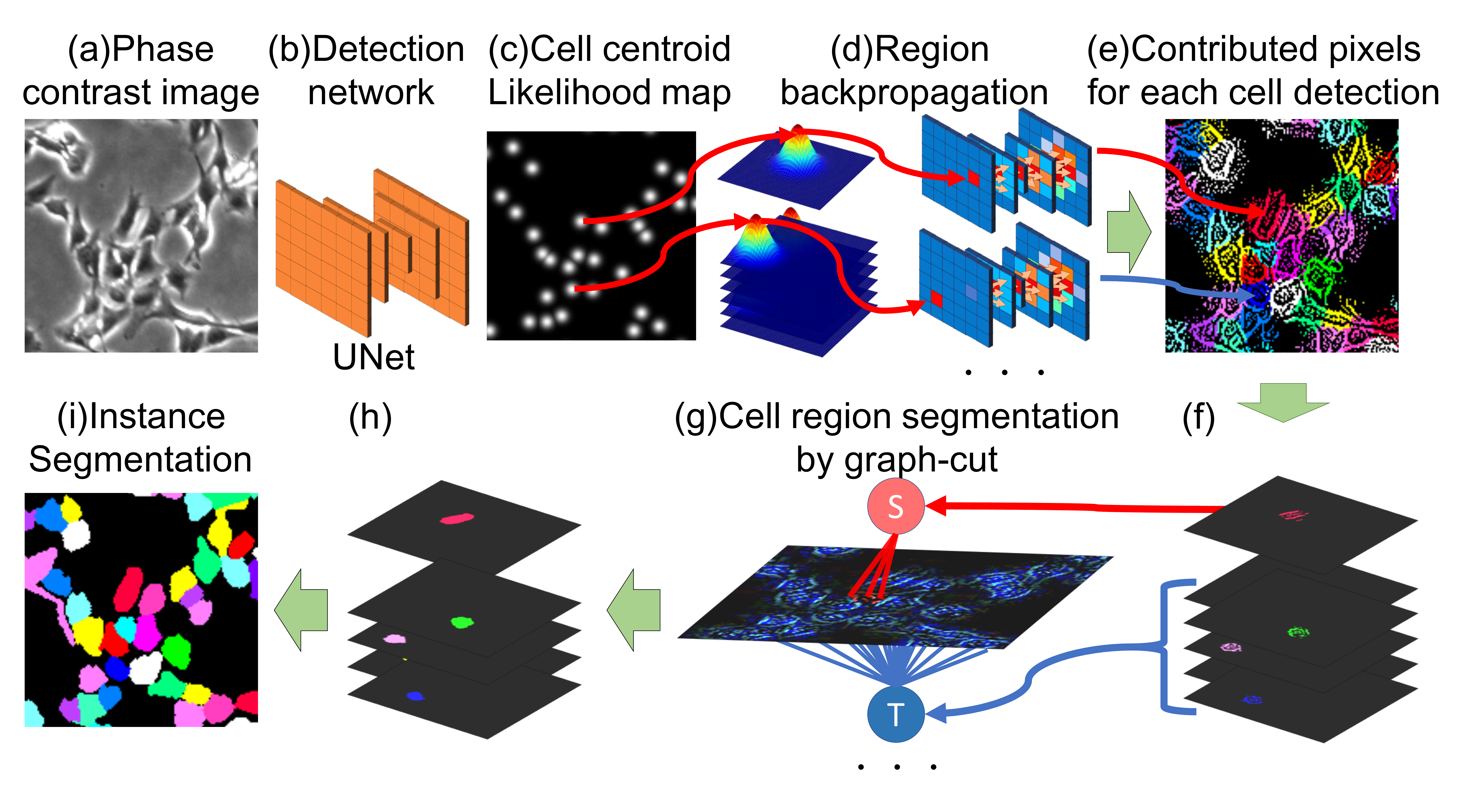}
    \caption{Outline of proposed method.}
    \label{fig:proposed}
\end{figure}

Noninvasive microscopy imaging techniques, such as phase contrast and differential interference contrast microscopy, have been widely used to capture cell populations for their appearance (shape) analysis and behavior analysis without staining.
Segmentation of individual cells is an essential task for such cell image analysis.
However, phase contrast microscopy images contain artifacts such as the halo and shade-off due to the optical principle as shown in Fig.~\ref{fig:proposed}(a). This makes segmentation difficult.
To address the difficulties, many CNN-based machine learning methods have been proposed.
In general, CNN requires a large amount of supervised data for each cell boundary.
Because cell shapes are complex, annotating individual cell shapes is very time-consuming. 
We should annotate cell regions under many situations such as types of cells, density, and microscopy.
In addition, the techniques for staining an entire cell region are not suitable for generation ground-truth of instance cell segmentation in dense conditions since it is difficult to separate the cell regions from such data.
We thus need to recognize cell shape from the simpler annotation.

Our aim is to develop a weakly supervised segmentation method that can segment individual cells in dense cell conditions by only using simple annotation, such as the centroid positions of individual cells, which does not include the cell shape (boundary) information.
The key assumptions of this study are as follows:

\begin{enumerate}
\item The rough centroid positions of individual cells (weak labels) are useful for an instance cell segmentation task that is difficult for direct segmentation methods without any training data; such weak label information makes segmentation easy. 
In addition, such rough centroid positions can be easily collected using fluorescent images in which the nuclei of cells are stained. 
\item In the process of deep learning for “detecting the center of individual cells”, the networks use "cell shape" information and thus the analysis of contributing pixels for detection is useful for the instance segmentation task.
\end{enumerate}
Based on the assumptions, we propose a weakly supervised instance cell segmentation method that first learns cell detection by CNN using centroid cell positions and segment individual cell regions by effectively using the contribution pixels for detection.
The main contributions of this paper are summarized as follows:
\begin{itemize}
\item To address the challenging task for segmenting individual cells, who touch each other with unclear boundaries, without the training data for cell regions, we propose a weakly supervised method that first detects the cell centroids using the weak label, and then uses the contribution pixel analysis in the trained detection network for instance cell segmentation.
\item For the contribution pixel analysis of detection, we use guided backpropagation~(GB)~\cite{gradcam}, which was developed to visualize pixels contributing to classification. GB backpropagates the responses from a class label node in the network. Instead, our method backpropagates the signals only from the particular regions of the output feature image in U-Net~\cite{ronneberger2015u}. Our method can extract the contributed pixels respectively.
\item 
We demonstrated the efficacy of our method using several microscopic images including multiple cell types captured by several types of microscopy techniques.
Our method achieved the highest accuracy compared with several conventional methods. In addition, we demonstrated that our method can perform without any annotation by using fluorescence images in which cell nuclei were stained as training data\footnote{In the test, the stained image is not required.}.
\end{itemize}
\section{Related work}
Many cell segmentation methods have been proposed for non-invasive microscopy images such as those taken with phase contrast microscopy using image processing methods based on intensity features~\cite{fogbank}, Graph-cut optimization~\cite{bensch}, and optical model-based method~\cite{Yin2012}.
These methods often do not work due to the differences in intensity distribution that arise from the difference in cell types (shape and thickness).
Many deep learning-based methods such as U-Net have been proposed.
However, these methods require sufficient training data that contains individual cell boundaries annotated by experts and such annotation is time consuming.

On the other hand, weakly supervised object segmentation methods for a general object have been proposed that use weak labels such as class labels and bounding boxes.
Zhou {\it et al.}~\cite{prm} proposed an instance segmentation method using only class labels as training data. This method implicitly supposes that objects to be segmented appear sparsely in an image. Therefore, it does not work for our target, in which many cells touch each other.
Li {\it et al.}~\cite{li2018weakly} uses faster R-CNN~\cite{ren2015faster} to detect the bounding boxes of the target objects, and segment individual object regions using a conditional random field (CRF), where they do not use network analysis. Their loss function is designed to exclude the region that the intersect regions of bounding boxes of nearby objects; this is a key for their method to succeed.
However, cells have a non-rigid and complex shape and they touch each other. Thus, the intersect regions of bounding boxes become much larger than those of general objects.

Unlike these methods, we effectively use a network analysis that can extract contributed pixels for each cell from a U-Net trained for cell detection in order to segment individual cell regions in dense conditions.

\section{Method}

Fig.~\ref{fig:proposed} shows an overview of the proposed method.
The proposed method first roughly detects the centroid positions of cells by a U-Net that is trained to output the likelihood map of cell positions (Fig.~\ref{fig:proposed}(c)). 
For each detected cell, the method performs back propagation from the center regions of a detected cell to extract the pixels that contributed to detecting the cell centroid(Fig.~\ref{fig:proposed}(d)).
Then, the information is used for graph-cut segmentation to segment individual cells, where the corresponding channel is used for the foreground seeds and the other channels are used for the background seeds(Fig.~\ref{fig:proposed}(g)).
\subsection{Cell detection with likelihood map}
A Bounding box has been widely used as a ground-truth of object detection.
As discussed above, however, a cell has non-rigid and complex shape and they often touch each other. In this case, a bounding box often contains multiple cell regions. Therefore, we use cell centroid positions as training data. This annotation is easier than annotating bounding boxes. 
The annotation may be off the cell centroid since the human annotations are not generally strict. To represent the gap between the human annotation and the true position, we use the cell position likelihood map as training data, where an annotated cell position becomes a peak and the value gradually decreases with a Gaussian distribution (Fig. ~\ref{fig:proposed} (c)).
To train the U-Net from the input of the original microscopy image, we use the mean of squared error loss function (MSE) between the estimated image by the U-Net and the ground-truth likelihood map. 
The output of U-Net defines $\bf{y}$.

\subsection{Propagating from detection map}
\begin{figure}[!t]
    \centering
    \includegraphics[width=\linewidth]{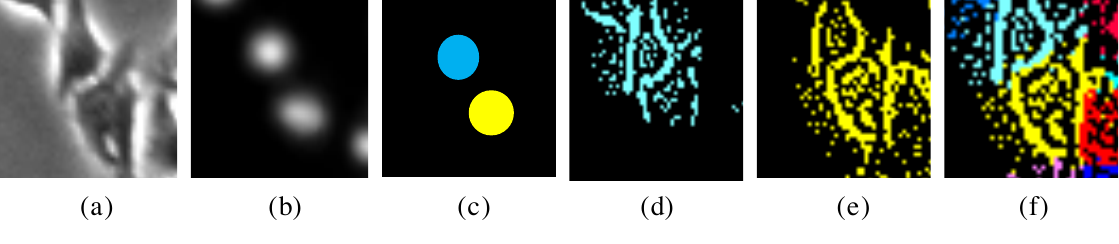}
    \caption{(a) Enlarged phase contrast image, (b) likelihood map, (c) each cell region, (d) and (e) each back propagation, (f) fused contribution map aggregating all cells in the image, where color indicates an individual cell.}
    \label{fig:one_props}
\end{figure}

The U-Net learns so that the cell center region has high values, and thus cell $u$'s center regions $\it{S}_u$ can be easily detected by thresholding the map and labeling the connected components as shown in Fig.\ref{fig:one_props}(c).

In the process of estimating the cell center region $\it{S}_u$ by U-Net, we consider that the other region in the cell also contributes to detection. Therefore, we apply Guided Backpropagation~\cite{gradcam} that visualizes the pixels contributed to classification.
As shown in Fig.~\ref{fig:proposed}(d), our method backpropagate the signals only from the particular regions of the likelihood map. Our method can extract the contributed pixels for each individual cells respectively.

The process of GB from the cell center regions $\it{S}_u$ is performed for all cells $u=1,...,N$, where $N$ is the number of cells.
We first initialize the response map $g^{Out}(u)$ so that the all regions outside the center regions $\it{S}_u$ substitute 0.
\begin{eqnarray}
    g_{i} ^ {Out}(u) &=&
    \begin{cases}
         y_i & if ~i\in{\it{S}_u}\\ 
         0    & if ~i\notin{\it{S}_u},
    \end{cases}
\end{eqnarray}
where $i$ denotes the pixel coordinates.
The GB is back propagating the signals from the output layer to the input layer using the trained parameters in the network. 
The GB is similar to compute the gradient of output to input.

A difference of normal gradient is the propagation at a ReLU function that uses both forward and backward pass information, where a forward pass in the estimation is recorded in the estimation process. 
We consider a case in which a ReLU function is at the $l+1$-th layer, where $g^{l+1}$ is the backward propagated value of the $l+1$-th layer, and $f^l$ is the forward propagated value of the $l$-th layer.
The signal is propagated to the $l$-th layer only if the both propagations are positive; otherwise, the backward propagation is to 0, formulated as:
\begin{eqnarray}
    g_{i} ^ {l}(u) &=& I(f^{l} _{i})\cdot I(g_{i}^{l} (u)) \cdot g_{i} ^ { l + 1 }(u)\\
    I(x), &=&
    \begin{cases}
        1&\mbox{if} ~x > 0\\
        0&\mbox{otherwise},
    \end{cases}
\end{eqnarray}
where $I$ is an indicator function.
Finally, the back propagated signal $g^{0}(u)$ is the map of the pixels that contributed to detecting the $u$-th cell.
For each detected cell, this process is respectively performed to obtain the contribution map of each cell.

However, the outside regions of the target cell also have values in the contribution map for each cell. 
Fig.~\ref{fig:one_props} shows the examples of the extracted contribution map from the touching cells. Fig.~\ref{fig:one_props}(d) is obtained from the blue region in Fig.~\ref{fig:one_props}(c), and Fig.~\ref{fig:one_props}(e) is obtained from the yellow region.
We can observe that the regions on the other cell also have positive values in each map.
That adversely affects the individual cell segmentation if we simply use this information for segmenting each cell.
To address this problem, we compared the values with the other channel ({\it i.e.}, cell) based on the basic of the fact that the pixel value $g_{i}^{0}(u)$ on the $u$-th cell in the $u$-th (cell's) contribution map tends to be larger than the values of the same pixel on the other cell's maps $g_{N}^{0} ~( \forall N \neq u)$.
The maximum projection contribution map $ {C}(u) $ for the $u$-th cell can be formalized as:
\begin{eqnarray}
    {C}_{i}  (u) &=& \phi_i(\argmax_{ k } g_{i}^0 ( k ), u).\\
    \phi_i(k ,u) &=&
    \begin{cases}
        {g}_{i}(k)&if ~(k=u) \\
        0&\mbox{otherwise}.
    \end{cases}
\end{eqnarray}\label{eq:indivisual region propagation}
This indicates that the value of the $i$-th pixel of the $u$-th map takes value only if the value is larger than the value of the same pixel $i$ on any other channels. This process is also done for each cell.
These maps are registered as channels  (Fig.~\ref{fig:proposed} (f)).
Fig.~\ref{fig:proposed}(e) shows the fusion image of all channels, where each color indicates an individual cell.

\section{Graph-cut}
We segment the individual cells independently by using graph-cut.
To segment the $u$-th cell, the proposed method uses the $u$-th contribution map ${C}(u)$ as a foreground seed and the maximum intensity projection image from all the other channels as a background seed. 
The saliency map of the original image is used for the data term. 
We simply use the inverse image of the original image for phase contrast microscopy images.
Then the final instance segmentation result is obtained by fusing all the segmented images as shown in Fig.~\ref{fig:proposed}(i).

\section{Experiments}

In the experiments, we evaluated the effectiveness of the proposed method compared with several methods in four challenging data-sets.
In the comparison, we selected three methods; Bensch ~\cite{bensch}, Chalfoun ~\cite{fogbank}, and Zhou ~\cite{prm}. 

The state-of-the-art method proposed by Li~\cite{li2018weakly}, which detects the bounding box using faster R-CNN, and then each object is segmented, is one of the most related works with our method. However, it totally did not work since the bounding boxes of nearby cells often overlap.
Instead, we selected the method proposed by Zhou~\cite{prm} that estimates the instance object regions by a deep neural network trained only using the class label.
To train their network, we prepared images to belong to the foreground (cell) that contains cells or background that does not contain by clipping the original images.
We prepared a small validation data for instance segmentation to tune the parameters of the methods of Bensch and Chalfoun since their methods require a small amount of annotation data about cell region boundaries as a validation data.

\begin{table}[t]
\begin{center}
\caption{Performance of compared methods.}\label{Table:performance}
    \scalebox{1}[1]{
        \begin{tabular}{cccccccc}
            \hline
            Metric & Data & & {\itshape Zhou\cite{prm}}& {\itshape Bensch\cite{bensch}}& {\itshape Yin\cite{Yin2012}} & {\itshape Chalfoun\cite{fogbank}}  &   Ours\\ \hline\hline
            \multirow{4}{*}{F-measure} & C2C12~\cite{Ker2018PhaseCT}& & 0.513 & 0.407 & 0.613 & 0.707 & \textbf{0.948} \\
            & NSC & &-& 0.645 &0.559 & 0.818 & \textbf{0.911} \\
            & B23P17 & &-& 0.209 & 0.719 & 0.679 & \textbf{0.887} \\
            & No annotation & &-& 0.669 & 0.403& 0.836 & \textbf{0.951} \\
            \hline
            \multirow{4}{*}{mDice} &
            C2C12~\cite{Ker2018PhaseCT} & & 0.244 &0.380 & 0.421 & 0.556 & \textbf{0.638} \\
            & NSC & & - & 0.167 &0.177 & 0.385 & \textbf{0.596} \\
            & B23P17 & & - & 0.061 & 0.487 & 0.354 & \textbf{0.598} \\
            & No annotation & & - & 0.165 & 0.343 & 0.500 &  \textbf{0.625} \\
            \hline\hline
        \end{tabular}
    }
\end{center}
\end{table}

\subsection{Dataset}
\begin{figure*}[t]
    \centering
      \includegraphics[width=\linewidth]{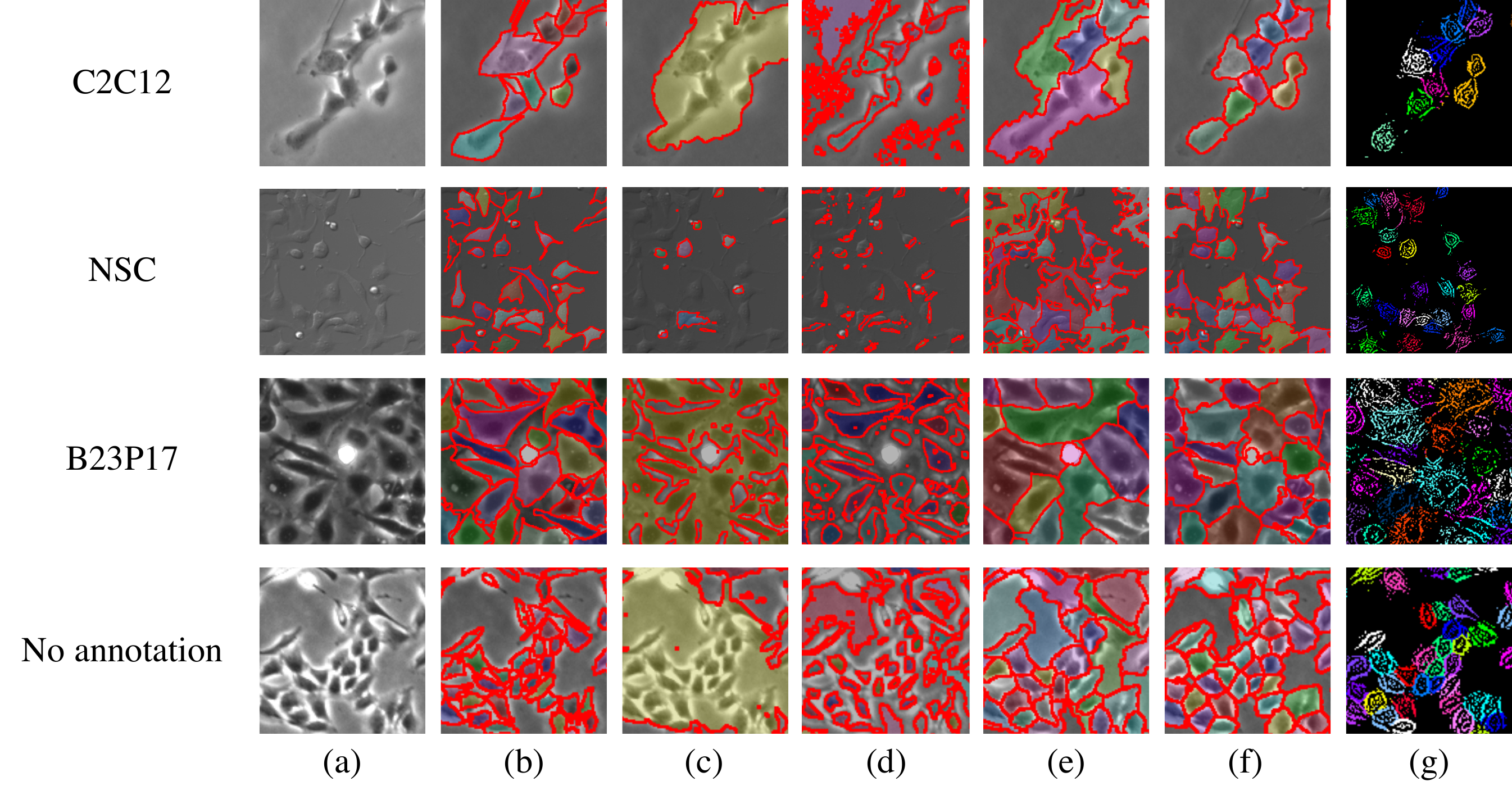}
     \caption{(a) Original image, (b) ground-truth,  (c) Bensch, (d) Yin, (e) Chalfoun, (f) ours, and (g) fused contribution map.}
    \label{fig:results}
\end{figure*}

We evaluated our methods by using three challenging data-set with annotation that cells were cultured in dense and captured by different microscopy as shown in Fig.~\ref{fig:results} (a); 1) C2C12: myoblast cells captured by phase contrast microscopy~\cite{Ker2018PhaseCT} at the resolution of $1040 \times 1392$ pixels, 2) B23P17: bovine epithelial cells captured by phase contrast microscopy at the resolution of $1040 \times 1392$ pixels,
3) NSC: neural stem cells captured by differential interference microscopy at the resolution of $512 \times 512$ pixels.
For ground-truth, cell centroids were roughly annotated in 880, 124, 120 images, respectively. In addition, the individual cell regions were annotated as the test data, where the total number of cells are 796, 711, and 416 cells. The images were divided into $320 \times 320$ patches to train our U-Net.

Furthermore, we evaluated our method using a dataset without any annotation; 4)~No annotation: myoblast cells captured by phase contrast microscopy with the fluorescence images, where cell nuclear were stained as training data, where the number of images is 86 pairs, and the resolution is $1360\times 1024$. For the test data, the boundaries of 1306 individual cells were annotated.

\subsection{Evaluation}
We first evaluated the performance of cell detection, where we use F-measure as the performance metric with the three-fold cross-validations.
Tab.~\ref{Table:performance} shows the average of the metrics for each data-set.
The proposed method outperformed the other methods in all data-set.
In particular, our method worked well from the dataset without human annotation ({\it i.e.}, instead, using fluorescence images).

Next, we evaluated the performance of the instance segmentation task.
We use the mean of Dice-coefficient (mDice) as the performance metrics.
To compute Dice, we first assigned the estimated cell regions and the regions of ground-truth.
The mDice is defined as $mDice =\frac{1}{M}\sum^{M}_{u=0}{\frac{2 \times tp^u}{2 \times tp^u + fn^u + fp^u}}$. $tp^u$ is the number of true positive pixels of cell $u$ (overlap regions), $fn^u$, $fp^u$ are the number of false negative and false positive pixels, respectively, where $M$ is the number of cells \footnote{In general, Dice takes a small value when the size of the object is small since the small discrepancy can affect to the metric. Since the size of a cell is much smaller than a general object, and thus it takes smaller value than that of a general object.}.

Tab.~\ref{Table:performance} summarizes the performance metrics\footnote{Zhou's method requires the training images that do not contain any cell. However, we could not make enough training data except (1) C2C12 due to the dense cells.}. In the results, our method outperformed other methods under all the dataset.
Fig.~\ref{fig:results} shows the examples of cell segmentation results of the compared methods.
Bensh's method, which uses the modified graph-cut, works for a cell that is located sparsely, but it did not work in dense conditions.
Yin's method, which uses the optical principle of phase contrast microscopy, did not work for cells that often change the thickness due to mitosis or differentiation.
Chalfoun's method, which first segments the cell cluster regions and then separates the individual cell regions, is sensitive to the intensity values, and thus it also did not work well.
Contrast to these methods, our method worked robustly under the various images since our method is based on the contribution map, where the contribution map showed good results Fig.~\ref{fig:results}.
In particular, our method produced good results even under the totally different images captured by different microscopy.
In addition, our method worked well using fluorescent images as training data. It demonstrated the possibility that our method does not require any human annotations anymore for any conditions.

\section{Conclusion}
We proposed a weakly supervised instance segmentation method that first detects the cell centroids using the weak label, and then uses the relevance pixel analysis in the trained detection network for instance cell segmentation.
It enables the challenging task for segmenting individual cell regions who touch each other with unclear boundaries in dense conditions without the training data for cell regions. 
Evaluation for instance segmentation tasks using several types of noninvasive microscopy image data-set demonstrated that the proposed method performs better than other methods.
We also demonstrated that instance segmentation is possible without annotations by using nuclei-stained images.
The very thin thickness regions around the boundaries of the cell were not correctly segmented by all the methods. It still remains challenges to segment such detailed regions in various conditions. This is our future work.
\\
\section*{Acknowledgement}
\vspace{-4mm}
This work was supported by JSPS KAKENHI Grant Number JP18H05104 and JP19K22895.
\vspace{5mm}

%

\bibliographystyle{splncs04}
\bibliography{refer}

\end{document}